\documentstyle[12pt]{article}
\newcommand{\beq}{\begin{equation}}
\newcommand{\eeq}{\end{equation}}
\newcommand{\beqa}{\begin{eqnarray}}
\newcommand{\eeqa}{\end{eqnarray}}
\newcommand{\ba}{\begin{array}}
\newcommand{\ea}{\end{array}}
\newcommand{\half}{\frac{1}{2}}

\begin{document}

\begin{flushright}
Preprint CAMTP/96-8\\
September 1996\\
\end{flushright}

\vskip 0.5 truecm
\begin{center}
\large
{\bf  Supersymmetric quantum mechanics based on higher excited states}\\
\vspace{0.25in}
\normalsize
Marko Robnik\footnote{e--mail: robnik@uni-mb.si}\\
\vspace{0.2in}
Center for Applied Mathematics and Theoretical Physics,\\
University of Maribor, Krekova 2, SLO--2000 Maribor, Slovenia\\
\end{center}

\vspace{0.3in}

\normalsize
\noindent
{\bf Abstract.} We generalize the formalism and the techniques of the
supersymmetric (susy) quantum mechanics to the cases where the superpotential
is generated/defined by higher excited eigenstates. The generalization
is technically almost straightforward but physically quite nontrivial since
it yields an infinity of new classes of susy-partner potentials,
whose spectra are exactly identical except for the lowest $m+1$
states, if the superpotential is defined in terms of the $(m+1)$-st
eigenfunction, with $m=0$ reserved for the ground state. It is
shown that in case of the infinite 1-dim potential well nothing new
emerges (the partner potential is still of P\"oschl-Teller
type I, for all $m$), whilst in case of the 1-dim harmonic oscillator
we get a new class of infinitely many partner potentials: 
for each $m$ the partner potential is expressed as the sum of the
quadratic harmonic potential plus rational function, defined as
the derivative of the ratio of two consecutive Hermite polynomials.
These partner potentials of course have $m$ singularities exactly
at the locations of the nodes of the generating $(m+1)$-st wavefunction.
The susy formalism applies everywhere between the singularities.
A systematic application of the formalism to other potentials
with known spectra
would yield an infinitely rich class of "solvable" potentials,
in terms of their partner potentials. If the potentials are
shape invariant they can be solved at least partially and new types of
analytically obtainable spectra are expected.
\vspace{0.6in}

Submitted to {\bf Journal of Physics A: Mathematical and General}
\normalsize
\vspace{0.1in}
  
\newpage

\section{Introduction}

After the classical papers of Witten (1981) and Gendenshtein (1983) 
the methods of
supersymmetric (susy)  (nonrelativistic) quantum mechanics have quickly
developed and it has been realized, that (1)  there exist partner
potentials with precisely the same energy spectra except for the
ground state ($m=0$) (whose wavefunction  $\phi(x)=\psi_0(x)$ is used to 
generate/define the superpotential   $W(x)$ - see below)
\footnote{the ground state energy $E_0^{(-)}$ is missing in the partner 
Hamiltonian $H_{+}$, so that its groundstate $E_{0}^{(+)}=E_{1}^{(-)}$}, 
and that (2) if they are {\em "shape invariant"}, their spectra 
and wavefunctions can be exactly and analytically solved. 
It is believed that the list of such shape
invariant partner potentials is now complete and finite
(L\'evai 1989, Barclay {\em et al} 1993), and therefore
quite limited in use. The research has been later further developed
also in direction of applying the WKB methods to such classes
of Hamiltonians, including the search for improved simple
quantization conditions which would be exact in case of susy
shape invariant potentials (Barclay, Khare and Sukhatme 1993,
Barclay and Maxwell 1991, Barclay 1993, Inomata, Junker and Suparmi
1993, Junker 1995, Robnik and Salasnich 1996), 
and also in direction of exploring the applicability of
the path integral techniques (Inomata and Junker 1991,1994). 
One of the nicest presentations
of susy quantum mechanics was published by Dutt, Khare and
Sukhatme (1988), henceforth referred to as DKS. We will use their notations. 
It should be mentioned at this place that
the ideas involved behind the susy property and shape invariance
were formulated first by Infeld and Hull (1951), where they were called
the "factorization method", and these authors refer further
to the related ideas in the works of Schr\"odinger (1940,1941).

\section{Generalized supersymmetric formalism}

The main point of this short Letter is to point out that the
whole formalism of the susy quantum mechanics can be generalized
to arbitrary higher excited eigenstates 
$\phi(x) = \psi_m(x), \quad m=0,1,2,\dots$,
used to generate the superpotential $W(x)$, namely

\beq
W(x) = - \frac{\hbar}{\sqrt{2\mu}} \frac{\phi'}{\phi},
\label{eq:susyW}
\eeq
where $\phi'(x) = d\phi/dx$, $\mu$ is the mass of the particle
moving in the $V_{-}$ potential, $2\pi \hbar$ is the Planck constant
and $m$ is the quantum number equal to the
number of nodes of the eigenfunctions $\psi_m(x)$ of the
starting potential  $V_{-}(x)$. The energy scale is adjusted so that
the $(m+1)$-st energy eigenvalue is exactly zero, $E_m^{(-)} =0$.
The corresponding Hamiltonian is
$H_{-} = -\frac{\hbar^2}{2\mu} \frac{d^2}{dx^2} + V_{-}(x)$, and the
 Schr\"odinger equation reads

\beq
H_{-} \psi_m^{(-)} = H_{-} \phi = (-\frac{\hbar^2}{2\mu} \frac{d^2}{dx^2}
+ V_{-}(x) )\phi = 0.
\label{eq:Schrodinger}
\eeq
Obviously, because $\phi'(x)\not = 0$ at the nodes $y_j$, 
the superpotential $W(x)$ will have
singularities at the nodes $y_j$, $j=1,2,\dots,m$ of $\phi$. 
However, this does not invalidate
our derivation, but it merely means, as will become clear later on, 
that the partner potential generated by $\phi$ diverges to $+\infty$
when $x\rightarrow y_j$, for any $j=1,2,\dots,m$. This implies that
the potential wells are well defined between two consecutive 
singularities and that they do not communicate with solutions in the
neighbouring wells. Thus if $m=0$ we have the common case of
usual susy potentials defined on $(-\infty,+\infty)$, 
if $m=1$ we have two separated potential
wells, each of them on a semiinfinite domain, for $m=2$ we
have one infinite potential well on a finite domain between two
nodes $y_1$ and $y_2$, and two binding potential wells on
the two seminfinite domains  $(-\infty,y_1]$ and $[y_2,+\infty)$,
and so on. The (partner) potentials constructed in this way
are nontrivial and certainly very interesting since they
contribute to our list of solvable potentials which now
becomes truly very rich and infinite in its contents.  
\\\\
In order to make this Letter selfcontained I will build up
the formalism necessary to construct the partner potentials
and to define the shape invariance, following DKS, in
order to demonstrate that the susy formalism does not break down
anywhere on its domain of definition, 
and to define the language needed to talk about
further results that I shall present in this contribution.
\\\\
First we express the starting potential $V_{-}(x)$ in terms of
the $(m+1)$-st eigenfunction  $\phi (x) = \psi_m(x)$, by
solving (\ref{eq:Schrodinger}) 

\beq
V_{-}(x) = \frac{\hbar^2}{2\mu} \frac{\phi''}{\phi},
\label{eq:V_minus}
\eeq
which is regular everywhere, because at the nodes $y_j$ the
second derivative $\phi''(x) = d^2\phi/dx^2$ also vanishes
with $\phi$. Thus the basic Hamiltonian $H_{-}$ reads

\beq
H_{-} = \frac{\hbar^2}{2\mu}(-\frac{d^2}{dx^2} + \frac{\phi''}{\phi}).
\label{eq:Hminus}
\eeq
The two important operators are:

\beq
A^{+} = \frac{\hbar}{\sqrt{2\mu}} (-\frac{d}{dx} - \frac{\phi'}{\phi}),
\label{operAplus}
\eeq
and

\beq
A = \frac{\hbar}{\sqrt{2\mu}} (\frac{d}{dx} - \frac{\phi'}{\phi}),
\label{operA}
\eeq
which gives

\beq
H_{-} = A^{+}A.
\label{HminusA}
\eeq
We further define the partner Hamiltonian $H_{+}$ and the partner
potential  $V_{+}$ as

\beq
H_{+} = A A^{+} = -\frac{\hbar^2}{2\mu} \frac{d^2}{dx^2} +V_{+}(x),
\label{eq:HplusA}
\eeq
where

\beq
V_{+}(x) = V_{-}(x) - \frac{\hbar^2}{\mu} \frac{d}{dx} (\frac{\phi'}{\phi})
\label{eq:Vplus}
\eeq
or

\beq
V_{+}(x) = -V_{-}(x) + \frac{\hbar^2}{\mu}(\frac{\phi'}{\phi})^2.
\label{eq:Vplusminus}
\eeq
The potentials $V_{+}$ and $V_{-}$ are called {\em susy-m partner 
potentials}.
We will show that they have the same energy levels, except for
the $(m+1)$ lowest states of $V_{-}$ for which there are no
corresponding states of $V_{+}$, so that the ground state of
the latter one is $E_0^{(+)} = E_{m+1}^{(-)}$. All higher states 
have then identical energies. From equation (\ref{eq:Vplusminus})
we see explicitly that at every node $y_j$, $j=1,2,\dots,m$ of
the defining eigenstate  $\phi=\psi_m^{(-)}$ the partner potential(s)
will have a singularity of the type $1/(x-y_j)^2$
such that $V_{+}(x)\rightarrow +\infty$
when $x\rightarrow y_j$, so that every branch of the partner
potential will be confining up to infinity, and the solutions
in various branches do not communicate. Thus for each $m$ we shall
find $(m+1)$ (branches of the) partner potentials.
\\\\
In terms of the superpotential $W$ defined in equation (\ref{eq:susyW})
we can write

\beq
\phi(x) = \psi_m^{(-)} (x) = \exp(-\frac{\sqrt{2\mu}}{\hbar} \int^x W(x)dx),
\label{eq:invW}
\eeq
which is well defined in the definition domain of any of the branches
of the partner potential, and obviously $\phi$ will go to zero where
$W$ has the poles $1/(x-y_j)$, as it should happen.
\\\\
Some of the relationships can be rewritten/reformulated in terms of
the superpotential $W(x)$ now:

\beqa
A^{+} & = &-\frac{\hbar}{\sqrt{2\mu}} \frac{d}{dx} + W(x),\nonumber\\
A & = & \frac{\hbar}{\sqrt{2\mu}} \frac{d}{dx} + W(x),\nonumber\\
\label{eq:AinW}
\eeqa
and also, the commutator of the operators $A$ and $A^{+}$ is

\beq
V_{\pm}(x) = W^2(x) \pm \frac{\hbar}{\sqrt{2\mu}} W'(x), \qquad 
W'(x)=\frac{dW}{dx}.
\label{eq:VinW}
\eeq
Further we observe 

\beq
V_{+} = V_{-} + \frac{2\hbar}{\sqrt{2\mu}} \frac{dW}{dx}.
\label{eq:VplustoVminus}
\eeq
and also

\beq
[A,A^{+}] = \frac{2\hbar}{\sqrt{2\mu}} \frac{dW}{dx}
\label{eq:commA}
\eeq
Now we have all tools at hand to show that the susy partner
potentials $V_{-}$ and $V_{+}$ are isospectral except for the
lowest $(m+1)$ states of $V_{-}$ which have no counterpart in
$V_{+}$, so that its ground state is  $E_0^{(+)} = E_{m+1}^{(-)}$.
\\\\
The demonstration, following DKS, is very easy: First we find that
if  $\psi_n^{(-)}$ is an eigenfunction of $H_{-}$ with the eigenenergy
$E_n^{(-)}$, then $A\psi_n^{(-)}$  is an eigenfunction of $H_{+}$
with the same energy:

\beq
H_{+}(A\psi_n^{(-)}) = AA^{+}A\psi_n^{(-)} = AH_{-}\psi_n^{(-)}=
AE_n^{(-)}\psi_n^{(-)} = E_n^{(-)} A\psi_n^{(-)}.      
\label{eq:isospec}
\eeq
Now we show that this applies only to the eigenstates $n$ higher
than $m$, $n=m+1,m+2, \dots$, by considering the normalization
condition, by writing the normalized state $\psi_n^{(+)} = 
C_nA\psi_n^{(-)}$, and calculating the normalizing coefficient $C_n$, 

\beq
\| \psi_n^{(+)} \| =  C_n^2 <A\psi_n^{(-)}|A\psi_n^{(-)}> =
 C_n^2 <\psi_n^{(-)}|A^{+}A\psi_n^{(-)}>  = C_n^2 
E_n^{(-)}\|\psi_n^{(-)}\|.
\label{eq:normalize}
\eeq
If all $\psi_n^{(-)}$ are normalized (they are certainly orthogonal,
because we deal with one dimensional systems, where degeneracies are
forbidden due to the Sturm-Liouville theorem (Courant and Hilbert 1968)
and therefore all eigenstates must be orthogonal),  then

\beq
C_n = \frac{1}{\sqrt{E_n^{(-)}}},
\label{eq:coeffC}
\eeq
which implies that the construction succeeds only iff $E_n^{(-)} > 0$,
implying that  $n>m$. Thus the two Hamiltonians $H_{-}$ and $H_{+}$
defined in (\ref{eq:Hminus}) and in (\ref{eq:HplusA}) are isospectral
except for the lowest $(m+1)$ eigenstates of $H_{-}$ which have no
counterpart in $H_{+}$.
\\\\
Counting now the eigenstates of $H_{+}$ from $n=0,1,2,\dots$, where
$n=0$ is the ground state, and $n$ is the number of nodes of the
(now also normalized) eigenfunction $\psi_n^{(+)}$, we have

\beq
\psi_n^{(+)} = \frac{1}{\sqrt{E_{m+1+n}^{(-)}}}A\psi_{m+1+n}^{(-)},
\qquad E_{n}^{(+)} = E_{m+1+n}^{(-)}.
\label{eq:normalizedpsi}
\eeq
Of course it is easy to show that, conversely, 
for every eigenstate $\psi_n^{(+)}$
of $H_{+}$ there exists the normalized eigenstate of $H_{-}$, namely

\beq
\psi_{m+1+n}^{(-)} = \frac{1}{\sqrt{E_n^{(+)}}} A^{+}\psi_n^{(+)},
\qquad n=0,1,2,\dots
\label{eq:psiminusofpsiplus}
\eeq
This completes our proof of isospectrality, generalized to the case
that the generating function $\phi$ of the superpotential $W$, defined in
equation (\ref{eq:susyW}), is a higher excited wavefunction, 
namely $\phi=\psi_m^{(-)}$, $m=0,1,2,\dots$. As we have seen, the
formalism of superpotential and of the partner potentials works
everywhere except at the singularities located at the nodal points
$y_i$ of $\phi$, where the partner potential $V_{+}$ goes to infinity
as $1/(x-y_i)^2$, thereby defining several branches of $V_{+}$ 
well defined on their disjoint domains of definition. 
\\\\
We have demonstrated that if one of the partner systems (the Hamiltonians) 
can be solved completely (by calculating the energy levels and the 
eigenfunctions), then the susy formalism enables one to solve the
partner problem completely, following equation (\ref{eq:normalizedpsi}).
One of the most important cases is of course the harmonic oscillator,
which we will discuss in detail below. 
\\\\
If the solutions for the two partner Hamiltonians are both unknown, 
then another approach is necessary to solve them. In case of the
standard susy formalism with $m=0$ we have the important class of
the shape invariant potentials. 
As is well known (DKS) the shape invariance of the two partner potentials
$V_{-}$ and $V_{+}$ is defined by

\beq
V_{+}(x; a_0) = V_{-}(x;a_1) + R(a_1),
\label{eq:shapeinvariance}
\eeq
where $a_0$ is a set of parameters, $a_1=f(a_0)$ and $R(a_1)$ is
independent of $x$. The procedure is now (essentially embodied in
the factorization method of Infeld and Hull (1951)) the following.
Consider a series of Hamiltonians  $H^{(s)}$, $s=0,1,2,\dots$,
where  $H^{(0)} = H_{-}$ and $H^{(1)} = H_{+}$, by definition

\beq
H^{(s)} = -\frac{\hbar^2}{2\mu} \frac{d^2}{dx^2} + V_{-}(x;a_s) 
+\sum_{k=1}^{s} R(a_k),
\label{eq:Hs}
\eeq
where 

\beq
a_s=f^s(a_0) = \underbrace{f\circ\dots\circ f}_{s} (a_0).
\label{eq:scompf}
\eeq
Now compare the spectra of $H^{(s)}$ with $H^{(s+1)}$, and 
find

\beqa
H^{(s+1)} & = & -\frac{\hbar^2}{2\mu} + V_{-}(x;a_{s+1}) + 
\sum_{k=1}^{s+1}R(a_k),\nonumber\\
H^{(s+1)} & = & -\frac{\hbar^2}{2\mu} + V_{+}(x;a_{s}) + 
\sum_{k=1}^{s}R(a_k),
\label{eq:partnerHs}
\eeqa
Thus it is obvious that $H^{(s)}$ and $H^{(s+1)}$ are susy partner
Hamiltonians, and they have the same spectra from the first level
upwards except for the ground state of $H^{(s)}$ whose energy
is

\beq
E_0^{(s)} = \sum_{k=1}^{s} R(a_k).
\label{eq:groundofHs}
\eeq
When going back from $s$ to $(s-1)$ we reach $H^{(1)}=H_{+}$ and
$H^{(0)} = H_{-}$, whose ground state energy is zero and its $n$-th
energy level being coincident with the ground state of the Hamiltonian
$H^{(n)}$, $n=1,2,\dots$. Therefore the complete spectrum of
$H_{-}$ is 

\beq
E_n^{(-)} = \sum_{k=1}^{n} R(a_k), \qquad E_0^{(-)} = 0.
\label{eq:finalspec}
\eeq

\vspace{0.2in}
\noindent
The generalization of shape invariance to the case of any $m\ge 0$ is 
straightforward, but it results in higher complexity and therefore
it is more rarely satisfied by the specific systems. By repeating the
above argumentation we reach the conclusion that, when 
(\ref{eq:shapeinvariance}) is satisfied for a superpotential $W$
with given $m$, then we cannot calculate the entire spectrum of
the shape invariant potential/Hamiltonian $H_{-}$, 
but only the subset (subsequence) of period $m+1$, namely

\beq
E_{m+n(m+1)}^{(-)} = \sum_{k=1}^n R(a_k), \qquad E_m^{(-)}=0, n=1,2,\dots.
\label{eq:mshapeinv}
\eeq
In the special case $m=0$ we of course recover the formula
(\ref{eq:finalspec}). For $m>0$ we have {\em none example} of
susy-m shape invariance so far.

\section{The example of the harmonic oscillator}

Let us consider a few examples of susy-m partner potentials, 
first the harmonic oscillator, defined by

\beq
V_{-}(x) = \frac{1}{2} \mu \omega^2 x^2 - (m+\frac{1}{2})\hbar\omega,
\label{eq:harmosc}
\eeq
shifted in energy so that 

\beq
E_m^{(-)} = 0.
\label{eq:mthlevel}
\eeq
Introducing the natural unit of length  $\alpha$ we can write
the ground state wavefunction as

\beq
\psi_0(x) = \frac{1}{\pi^{1/4}\alpha^{1/2}} \exp(-\frac{x^2}{2\alpha^2}),
\qquad \alpha = \sqrt{\frac{\hbar}{\mu \omega}},
\label{eq:groundst}
\eeq
Defining the creation operator $a^{+}$,

\beq
a^{+} = \frac{1}{\sqrt{2}}(-\alpha \frac{d}{dx} + \frac{x}{\alpha}),
\label{eq:aplus}
\eeq
we can write down all the eigenfunctions, in particular the $(m+1)$-st
one, labelled by $m$, and denoted by $\phi$, as follows

\beq
\psi_{m}^{-}(x) = \frac{(a^{+})^{m}}{\sqrt{m!}} \psi_0(x) = \phi(x).
\label{eq: mthwavefunction}
\eeq
Now we calculate the superpotential according to equation (\ref{eq:susyW}),
by using the following operator when calculating $\phi'=d\phi/dx$,
obtained from equation (\ref{eq:aplus}),

\beq
\frac{d}{dx} = \frac{1}{\alpha}(\frac{x}{\alpha} - \sqrt{2} a^{+}),
\label{eq:ddx}
\eeq
and find

\beq
W(x) = -\frac{\hbar}{\alpha\sqrt{2\mu}}\left( \frac{x}{\alpha} - \sqrt{2(m+1)}
\frac{\psi_{m+1}(x)}{\psi_m(x)}\right).
\label{eq:susyWharm}
\eeq
Using the explicit solution for $\psi_m(x)$, namely

\beq
\psi_m(x) = 
\left( \frac{1}{2^mm!}\sqrt{\frac{1}{\pi\alpha^2}} \right)^{1/2}
{\rm H}_m(\frac{x}{\alpha}) \exp(-\frac{x^2}{2\alpha^2}),
\label{eq:mthwavefunction}
\eeq
where ${\rm H}_m(z)$ is the Hermite polynomial (Abramowitz and Stegun 1965),
we obtain

\beq
\frac{\psi_{m+1}(x)}{\psi_m(x)} = \frac{1}{\sqrt{2(m+1)}} 
\frac{{\rm H}_{m+1}(\frac{x}{\alpha})} {{\rm H}_{m}(\frac{x}{\alpha})} 
\label{eq:psiratio}
\eeq
and hence

\beq
W(x) = -\frac{\hbar}{\alpha\sqrt{2\mu}} \left( \frac{x}{\alpha} - 
\frac{{\rm H}_{m+1}(x/\alpha)}{{\rm H}_{m} (x/\alpha)} \right).
\label{eq:susyWharmf}
\eeq
From this and equation (\ref{eq:VplustoVminus}) we get finally
the susy-m partner potential of the harmonic potential
(\ref{eq:harmosc}), namely

\beq
V_{+}(x) = \half\mu\omega^2x^2 - (m+\frac{3}{2})\hbar\omega
+ \hbar\omega \frac{d}{dz} \left( \frac{{\rm H}_{m+1}(z)}{{\rm H}_{m}(z)} 
\right)_{z=\frac{x}{\alpha}}
\label{eq:Vplusharm}
\eeq
whose energy levels are the same as for the (energy shifted) 
harmonic oscillator, namely

\beq
E_n^{(+)}= (n+1)\hbar\omega, \qquad n=0,1,2,\dots
\label{eq:partnerspec}
\eeq
so that the ground state energy $E_0^{(+)} = \hbar\omega$, which is
equal to the $(m+1)$-st energy level of (\ref{eq:harmosc}).
\\\\
This result is important, because for $m>0$ it yields new 
interesting potentials with purely discrete spectrum, isospectral
to the harmonic oscillator, except for the lowest $(m+1)$-st 
eigenstates. Let us look just at the few lowest cases.

\beqa
& {\bf m=0:} & \quad V_{+}(x)= \half\mu\omega^2x^2 + 
\half\hbar\omega,\nonumber\\
& {\bf m=1:} & \quad V_{+}(x)= \half\mu\omega^2x^2 - \half\hbar\omega
     + \hbar\omega\frac{\alpha^2}{x^2},\nonumber\\
& {\bf m=2:} & \quad V_{+}(x)= \half\mu\omega^2x^2 - \frac{3}{2}\hbar\omega
+ 4\hbar\omega \left( \frac{1}{2z^2-1} +\frac{2}{(2z^2-1)^2} \right),
\; z=\frac{x}{\alpha}, \nonumber\\
& {\bf m=3:} & \quad V_{+}(x)= \half\mu\omega^2x^2 -\frac{5}{2}\hbar\omega
     + \frac{3\hbar\omega}{z^2} \left( 1 + \frac{6}{2z^2-3}+
\frac{12}{(2z^2-3)^2} \right), \; z=\frac{x}{\alpha}. \nonumber\\
\label{eq:mcases}
\eeqa
The case $m=0$ is the usual susy-0 case showing just that the 1-dim harmonic
oscillator potential is indeed susy-0 shape invariant. But $m=1$
gives a new example, which nevertheless is well known as the radial
problem of the 3-dim harmonic potential, which is thus susy-1 partner
potential of the 1-dim harmonic oscillator potential. It has only
one singularity at $x=0$. Next, for $m=2$
in the above list, we see the first new nontrivial example, of a specific
rational potential which is susy-2 partner potential of the 1-dim
harmonic potential. It has singularities at the two nodes
$x=y_1=-\alpha/\sqrt{2}$ and $x=y_2=+\alpha/\sqrt{2}$, of the type 
$1/(x-y_j)^2$. Therefore it has three branches (ranges),
namely  $(-\infty,y_1]$, $[y_1,y_2]$, and $[y_2,+\infty)$.
The spectrum is identical in each of them. Further, in case $m=3$, we have
three singularities at the nodes where $x$ is equal to 
$y_1=-\alpha\sqrt{3/2}$, $y_2=0$, and $y_3=-y_1=+\alpha\sqrt{3/2}$,
and thus we have two independent different potentials within the two
ranges $[y_2,y_3]$ and $[y_3,+\infty)$. (The other two ranges confine
the potentials which are equivalent due to the evenness of $V_{+}(x)$.)
\\\\
For higher $m$ we get
new classes of interesting rational potentials, all of them
being isospectral to the harmonic oscillator except for the
lowest $m+1$  eigenstates of the latter which are missing in
the partner potentials. For each $m$ we have rational potentials
with $m+1$ branches, defined by the $m$ nodes $y_j$, $j=1,\dots,m$.
Since the Hermite polynomials ${\rm H}_m(z)$ are even or odd functions of $z$, 
depending on whether $m$ is even or odd, the superpotential
$(\ref{eq:susyWharmf})$ is always odd function of $x$ and therefore
the susy-m partner potential  $V_{+}(x)$ in equation (\ref{eq:Vplusharm})
is always even function of $x$ with $m$ singularities.
\\\\
Asymptotically when
$|x| \rightarrow \infty$ the potential behaves still as harmonic
quadratic potential with the leading term $\half \mu\omega^2x^2$,
which is true for any $m$, as can be shown using the asymptotic 
properties of the Hermite polynomials, as described below.
\\\\
The limiting (semiclassical) behaviour of the potential $V_{+}$ 
when $\hbar\rightarrow 0$ is interesting. It implies that $z=x/\alpha=
x\sqrt{\mu\omega/\hbar}$ tends to $+\infty$  and therefore from
the asymptotic properties of the Hermite polynomials 
${\rm H}_{m+1}(z)/{\rm H}_{m}(z) \rightarrow 2z$, for $z\rightarrow 
+\infty$, we conclude

\beq
V_{+}(x) \longrightarrow \half\mu\omega^2 x^2 - (m-\half)\hbar\omega
\longrightarrow  \half\mu\omega^2x^2 \quad {\rm when} \; 
\hbar\rightarrow 0.
\label{eq:asymptoticVplus}
\eeq
Thus the semiclassical limiting form of all these potentials
is just the harmonic quadratic potential, meaning that all the rational
potentials in (\ref{eq:Vplusharm})  
all have zero classical limit. From (\ref{eq:Vplusharm}) 
it is clear that the harmonic oscillator potential is {\em not}
susy-m shape invariant, except for $m=0$, which is the familiar
case of shape invariance (see DKS).

\section{Discussion and conclusions}

Using the same formalism applied to known solvable potentials 
for various $m$ we can systematically
construct the vast class of new potentials which will be isospectral 
to each of the known solvable potentials, almost all of them 
being susy-0 shape invariant, and listed in DKS. 
\\\\
Finally we can state the result which can be easily  verified
in a straightforward manner (we omit the derivation due to
the lack of space here), that the P\"oschl-Teller type I
potential is the susy-m partner potential of the infinite 
potentiall well {\em for any value of $m\ge0$}. 
At present we do not know any
specific cases of susy-m shape invariance with $m>0$,
and also have no further calculations for susy-m partner
and solvable potentials, which remains as the future project.

\section*{Acknowledgements}
\par
The financial support by the Ministry of Science
and Technology of the Republic of Slovenia is acknowledged with
thanks.

\newpage

\section*{References} 
\parindent=0. pt
Abramowitz M and Stegun I 1965 {\it Handbook of Mathematical Functions}
(New York: Dover)
\\\\
Barclay D T 1993 {\it Preprint} UICHEP-TH/93-16 Nov 1993
\\\\
Barclay D T, Dutt R, Gangopadhyaya A, Khare A, Pagnamenta A and Sukhatme U
1993 {\it Phys. rev. A} {\bf 48} 2786
\\\\
Barclay D T, Khare A and Sukhatme U 1993 {\it Preprint} UICHEP-TH/93-13
Sept 1993
\\\\
Barclay D T and Maxwell C J 1991 {\it Phys. Lett. A} {\bf 157} 357
\\\\
Dutt R, Khare A and Sukhatme U P 1988 {\it Am. J. Phys.} {\bf 56} 163
(referred to in text as DKS)
\\\\
Gendenshtein L E 1983 {\it Pisma Zh. Eksp. Teor. Fiz.} {\bf 38} 299
(English translation in {\it JETP Lett.} {\bf 38} (1983) 356)
\\\\
Infeld L and Hull T E 1951 {\it Rev. Mod. Phys} {\bf 23} 21
\\\\
Inomata A and Junker G 1993 in {\it Lectures in Path Integration: Trieste
1991} eds. H. A. Cerdeira {\em et al} (Singapore: World Scientific)
\\\\
Inomata A and Junker G 1994 {\it Phys. Rev. A} {\bf 50} 3638
\\\\
Inomata A, Junker G and Suparmi A 1993 {\it J. Phys. A: Math. Gen.} {\bf 
26} 2261
\\\\
Junker G 1995 {\it Turk. J. Phys.} {\bf 19} 230
\\\\
L\'evai G 1989 {\it J. Phys. A: Math. Gen.} {\bf 22} 689 
\\\\
Robnik M and Salasnich L 1996 {\it Preprint} CAMTP/96-5 July 1996
\\\\
Schr\"odinger E 1940 {\it Proc. R. Irish Acad.} {\bf A46} 9
\\\\
Schr\"odinger E 1941 {\it Proc. R. Irish Acad.} {\bf A46} 183
\\\\
Witten E 1981 {\it Nucl. Phys. B} {\bf 185} 513
\\\\
\end{document}